\documentclass[11pt,notitlepage,longbibliography]{revtex4-2} 
\usepackage{amsmath,amssymb,mathtools,mathrsfs}
\usepackage[colorlinks=true,allcolors=black]{hyperref} % Replaces color3links
\usepackage[usenames,dvipsnames]{xcolor}
\usepackage{xcolor}
\usepackage{ulem}
\usepackage[a4paper,textwidth=18.5cm]{geometry}
\usepackage{soul} % for strikethrough \st
\usepackage{siunitx}
\DeclareSIUnit{\litre}{l}
\usepackage{changes}
\begin{document}

\title[Article Title]{Dynamics of Thin Lubricant Films upon Liquid Contact on Slippery Surfaces}

\author{
	Shivam Gupta\textsuperscript{1}, 
	Bidisha Bhatt\textsuperscript{1}, 
	Zhaohe Dai\textsuperscript{2,*}, and
	Krishnacharya Khare\textsuperscript{1,*} \vspace{1em}
}

\affiliation{\textsuperscript{1}Department of Physics, Indian Institute of Technology Kanpur, Kanpur 208016, Uttar Pradesh, India \\ \textsuperscript{2}Department of Mechanics and Engineering Science, State Key Laboratory for Turbulence  and \\ Complex Systems, College of Engineering, Peking University, Beijing 100871, China}

\begin{abstract}
In recent years, slippery surfaces have attracted significant interest due to their excellent liquid-repellent properties and their potential in diverse commercial applications. Such surfaces are prepared by coating functionalized solid substrates with a thin lubricant film that prevents direct contact between a liquid and the substrate. The morphology of thin films upon liquid contact plays a central role in governing various phenomena, including the coalescence and mobility of liquid droplets, heat transfer efficiency, and  the extent of lubricant depletion. However, a detailed understanding of film dynamics upon droplet contact remains limited, both from theoretical and experimental perspectives. Here, by employing principles of fluid dynamics, optics, and surface wetting, we present a comprehensive study that examines both the spatial and temporal variations of lubricant films upon contact with sessile liquid droplets and liquid bridges. Our findings reveal that the film dynamics can be categorized into three distinct stages, each significantly influenced by key system parameters: initial film thickness, three-phase contact line width, and Laplace pressure of liquids. Furthermore, we demonstrate that by optimizing these parameters, it is possible to reverse the lubricant flow in the final stage, thereby causing the liquid to partially lift off from the slippery surface. 
\end{abstract}

\maketitle

\vspace{5em}

\vfill
\begin{flushleft}
	\rule{\textwidth}{0.8pt}  % Thick horizontal line (full width, 0.8pt height)
	\small\textit{email: \href{mailto:daizh@pku.edu.cn, kcharya@iitk.ac.in}{daizh@pku.edu.cn, kcharya@iitk.ac.in 
	}}
\end{flushleft}

\newpage
%\section{Introduction}\label{sec1}

Inspired by \textit{Nepenthes} pitcher plants, the strategy of coating surfaces with suitable thin lubricant films has emerged as a powerful method to prevent direct contact between a liquid and the solid \cite{nepenthes_bohn_2004,nepenthes_2009,wong2011}. Previous studies have demonstrated that when the lubricant films are stable, liquid droplets move easily on lubricated surfaces, a characteristic often described as slippery. Slippery surfaces have shown potential in a wide range of applications, including self-healing, anti-icing, water harvesting, heat transfer, and anti-biofouling \cite{comparison_of_shs_and_slippery_2020,wong2011,antiicing_aizenberg_2, antiicing_aizenberg,water_harvesting_2018,heat_transfer_varanasi_2012,antibiofouling_aizenberg}.
 
When a liquid droplet contacts a slippery surface, unbalanced forces at the three-phase contact line draw the lubricant upward, forming a negative pressure region, known as the wetting ridge. The pressure difference drives the lubricant flow toward the ridge from both beneath and beyond the droplet, resulting in a time-dependent evolution of the lubricant morphology. Studies have shown that thicker films beneath liquid and larger wetting ridges enhance droplet mobility and coalescence, respectively, whereas thinner films beneath liquid and smaller ridges are more effective in improving heat transfer efficiency and reducing lubricant depletion, respectively  \cite{ lubricantdepletion,vollmer_lubricant_depletion_2021,KrederPRX, Non_wettable_Surfaces_book_2016, drop_coalescence_1, drop_coalescence_2, Smith, dropfriction}. Despite these insights, a detailed understanding of lubricant dynamics both beneath and beyond a sessile liquid droplet, remains limited. This largely stems from several experimental challenges, including difficulty in observing thin-films due to poor refractive index contrast, unstable lubricant films, and moving contact lines due to droplet evaporation and mobility.

\begin{figure}[h]
    \includegraphics[width=17.8cm]{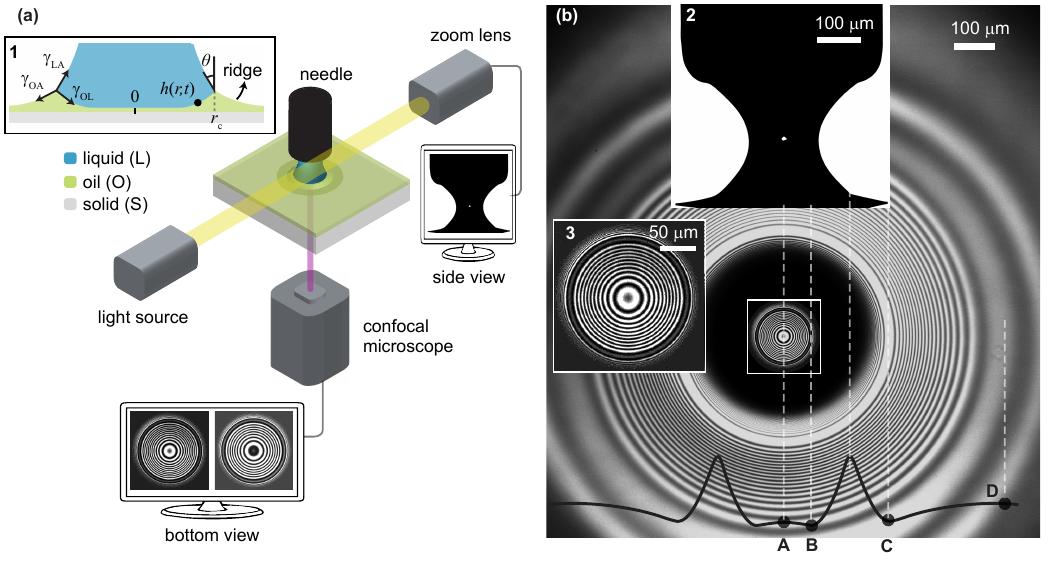}
    \caption{
\textbf{Visualization of system.} (a) Schematic of the experimental setup used to observe the system simultaneously from side and bottom. Inset 1 shows the schematic of the system depicting various parameters.
Surface tensions of the oil-air, oil-liquid, and liquid-air interface are represented by $\gamma_\mathrm{OA},\gamma_\mathrm{OL}$, and $\gamma_\mathrm{LA}$, respectively.  $h(r,t)$ is the thickness of the oil film. (b) Aligned side-view (inset 2) and bottom-view interferometry images of the system captured 1 minute after Krytox, held by a needle, contacted a silicone oil  coated glass surface (initial film thickness $h_0$ = \SI{5.9}{\micro\metre}). The interference pattern (captured using a red laser with wavelength $638$ nm) is due to the thin silicone oil film sandwiched between Krytox and glass. Brightness of the pattern beneath Krytox is enhanced for better visibility and the magnified view is shown in inset 3. Spatial profile of the oil (not to scale) along the mid-section is displayed below, highlighting the film's local maxima and minima both beneath and beyond the liquid. The features are defined as A-center, B-edge, C-neck, and D-bump. }
    \label{setup}
\end{figure}

We addressed the challenges of liquid evaporation and poor refractive index contrast by using an unconventional system of Krytox (a perfluorinated oil), silicone oil, and glass as the top liquid, lubricant and solid, respectively, and imaged the system with a confocal microscope. To unambiguously resolve film thickness, the microscope was configured with two lasers ($\lambda_1=488$ nm and $\lambda_2=638$ nm) and two separate detectors, operating in a dual-channel mode (Methods and Supplementary Section 3). To ensure that the system satisfied the lubricant stability criteria, the glass surfaces were chemically modified with polymer brushes (Methods and  Supplementary Sections 4 and 5) \cite{ Lafuma_2011, Smith, oleoplaning}. The unintended droplet motion was prevented by holding one end of the needle using a needle (inset 2 of Fig. \ref{setup}). It was found that for the complete immobilization of Krytox, the needle size should be comparable to the width of the three-phase contact line.  We thus study the thin-film dynamics upon contact with positive Laplace pressure liquid bridges (hereinafter referred to as liquid bridges) and provide insights into the dynamics for the case of sessile droplets.

Due to the low contact angle of Krytox on silicone oil ($\approx 40^\circ$), Krytox, held at one end by a needle, contacts the surface with a negative Laplace pressure. Thus, subsequent to contact, the needle was retracted to a set height, forming a positive Laplace pressure liquid bridge. To simultaneously determine the Laplace pressure of liquid bridges and the profile of thin-films, the system was visualized from the side and bottom. Fig. \ref{setup} shows the employed experimental setup along with a typical side and bottom view of the system.  We define the endpoints of the wetting ridge, located beneath and beyond the liquid, as edge and neck, respectively, and the first local maxima of film beyond the liquid as bump (Fig. \ref{setup}(b)).

\clearpage

\clearpage
\section{Dynamics beneath sessile liquids}\label{}
\begin{figure}[h]
	\centering
	\includegraphics[width=17.8cm]{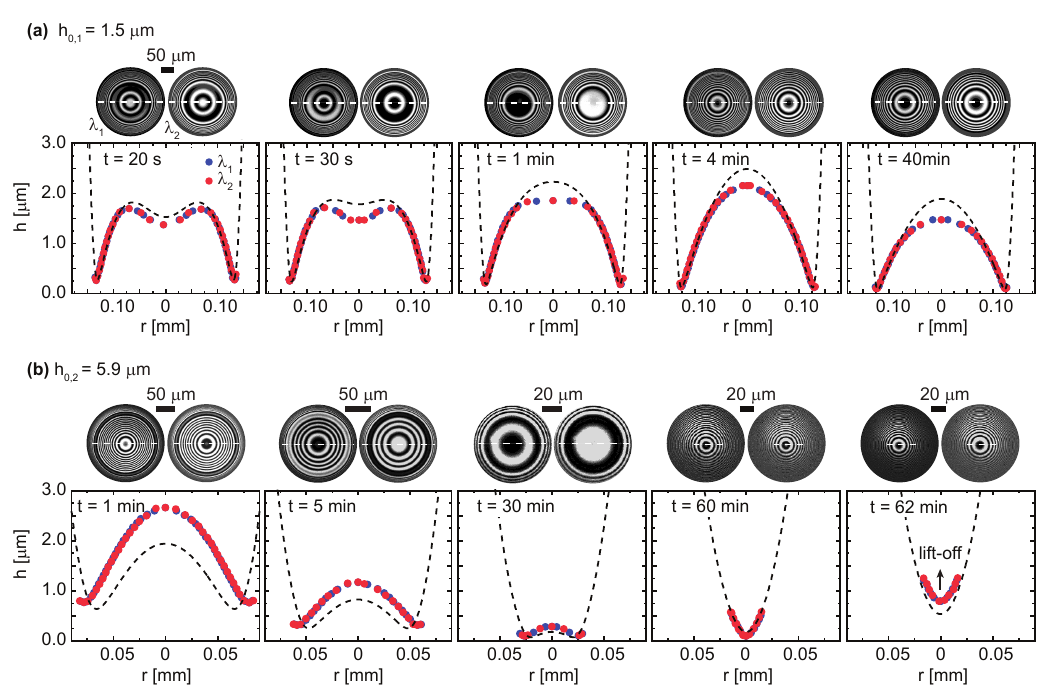}
	\caption{\textbf{Spatial variation of lubricant film beneath sessile liquid.} Interferometric images captured over time for a silicone oil film beneath sessile Krytox with initial film thicknesses (a) $h_{0,1}=1.5 \,$\SI{}{\micro\metre} and (b) $h_{0,2}=5.9 \,$\SI{}{\micro\metre}  with corresponding determined profile along the midsection (depicted by white dashed lines) displayed in the bottom panels. The images are captured simultaneously using blue ($\lambda_1=488$ nm) and red laser ($\lambda_2=638$ nm) and the profiles are reconstructed from the intensity information  (Supplementary Section 3 for more details). Blue and red filled circles correspond to the film height determined using blue and red lasers, respectively. The black dashed lines represent the numerical solutions. Film profile for the case of $h_{0,1}$ transitions from a wimple to a dimple, while no such transition is observed for the case of $h_{0,2}$. Instead, the edge merge and leads to liquid lift-off for the latter. 
	}
	\label{spatial_exp}
\end{figure}

To begin with, we experimentally determined the spatial evolution of the thin-film beneath liquid bridges over time scales ranging from tens of seconds to tens of minutes. Fig. \ref{spatial_exp} displays the interferometry images, along with the corresponding film profiles shown in the bottom panels for two different initial thicknesses,  $h_{0,1}= 1.5\SI{}{\,\micro\metre}$, and $h_{0,2}= 5.9\SI{}{\,\micro\metre}$, with similar values of contact line radius $r_\mathrm{c}$ and Laplace pressure of liquid bridge $\Delta P_\mathrm{liq}$. 
For the case of $h_{0,1}$, the thin-film initially exhibited a wimple (characterized by a central minimum, followed by maxima, and then another minima), which gradually transitioned into a dimple (characterized by a central maximum and followed by minima) within about 4 min. During this transition, the height at the center increased to about $2.1 $\SI{}{\,\micro\metre}, exceeding the initial thickness $h_{0,1}$. The transition behavior from wimple to dimple is similar to the thin-film dynamics when liquids within an external ambient liquid are pushed from very close to a wall  \cite{increasethick}. As time progressed, the lubricant continued to drain beneath the liquid bridge while retaining the dimple profile. 
For the case of $h_{0,2}$, we observed a slightly different behavior compared to the case of $h_{0,1}$. Initially, the lubricant did not exhibit a wimple; instead, the thin-film dynamics began with a dimple profile and retained this profile during most of the observation period.     The beneath lubricant film also exhibited a significantly faster drainage rate as compared to $h_{0,1}$. During the drainage process, the edge continued to propagate toward the center and, remarkably, merged after about $58$ min. Subsequently, the liquid bridge partially lifted off from the slippery surface (mediated by the lubricant), and the height at the center began to increase.
Notably, this increase occurred at a significantly faster rate, with height at the center reaching about $800$ nm within just 4 minutes of merging. To our knowledge, the lift-off (hereinafter referred to as lift-off) behavior of sessile liquids on slippery surfaces has not been reported previously.
To further investigate the lift-off phenomenon, we conducted additional set of experiments for similar values of $r_\mathrm{c}\sim  150 \SI{}{\,\micro\metre}$. In experiments with thinner films ($h_0\sim 1 $\SI{}{\,\micro\metre}), we did not observe lift-off, whereas in experiments with thicker films ($h_0\sim 6 $\SI{}{\,\micro\metre}), we consistently observed lift-off. Additionally, we found that for a fixed initial thickness, the lift-off time decreases with decrease in the width of three-phase contact line. 

In addition to the faster drainage rate for the case of $h_{0,2}$, we also noted a significant increase in the size of wetting ridge, which led to a decrease in the values of $r_\mathrm{c}$ (from about 155 to 141 $\SI{}{\,\micro\metre}$) and $\Delta  P_\mathrm{liq}$ (from about 44 to 36\,$\SI{}{N/m^2})$ over time. For the case of $h_{0,1}$, however, we did not observe a significant change in the values of $r_\mathrm{c}$ ($\sim$ 153 $\SI{}{\,\micro\metre}$) and $\Delta  P_\mathrm{liq}$ ($\sim$ 45\,$\SI{}{N/m^2})$ (Supplementary Section 7).
Therefore, the experimental observations collectively indicate that thin-film  dynamics beneath sessile liquids evolve through multiple stages, are strongly influenced by the initial film thickness and contact line width, and can lead to liquid lift-off in specific cases. 

\begin{figure}[h]
    \centering
    \includegraphics[width=17.8cm]{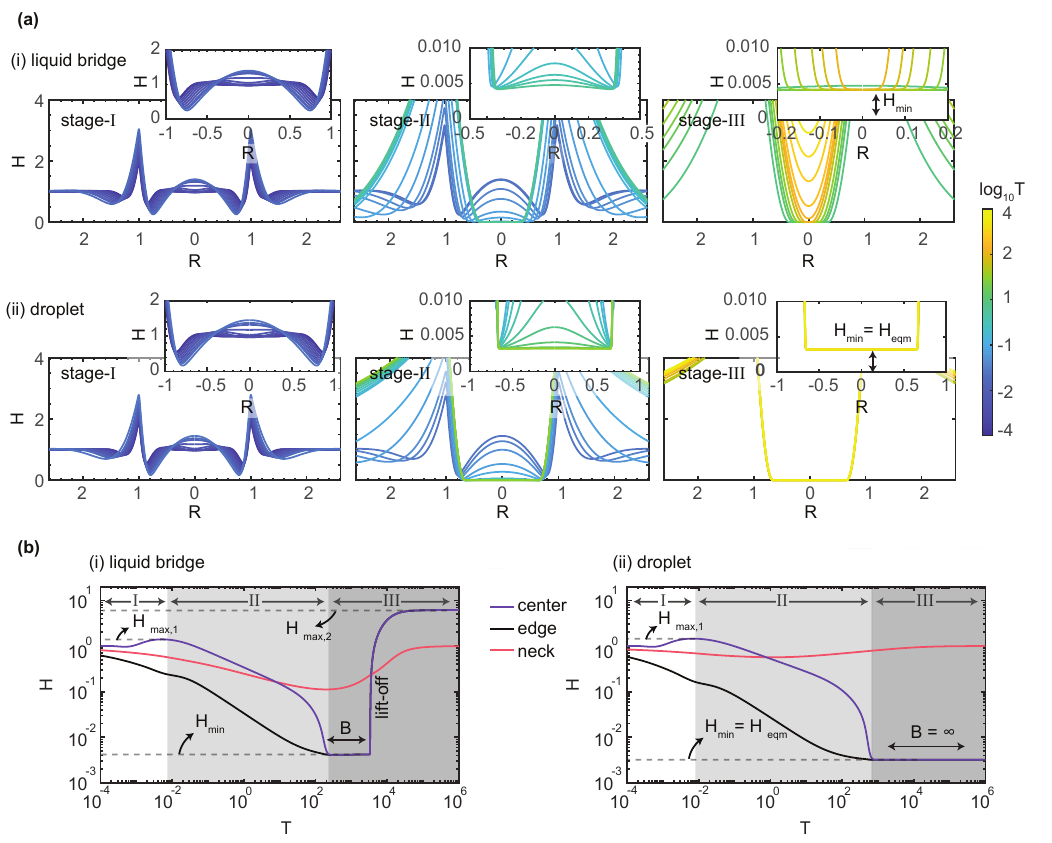}
    \caption{\textbf{Lubricant film dynamics: numerical solutions.} (a) Spatial variation of dimensionless film height $H$ with dimensionless position $R$ upon contact with (i) a liquid bridge and (ii) a sessile droplet. The liquid is present in the region $0\le R\le 1$. 
    The change in dimensionless time $T$ is depicted with different colors with each color representing a different time as indicated by the colorbar on the right. The corresponding magnified view of the film profile beneath the liquid is displayed in the top panels. (b) Temporal variation of $H$ at center, edge and neck with dimensionless time $T$. 
    Lubricant dynamics progress through three stages: I, II, and III. The thickness at the center reaches a maximum value $H_\mathrm{max,1}$ in stage-I and      decreases to a minimum thickness $H_\mathrm{min}$ within the van der Waals range in stage-II. Dynamics in stage-III are different for the case of liquid bridge and sessile droplet. For a liquid bridge, depending on the system parameters, merging of edge may occur in stage-III, resulting in either 
    $H_\mathrm{min} = H_\mathrm{eqm}$  if the edge do not merge, or  $H_\mathrm{min} < H_\mathrm{eqm}$  if they do. If the edge merge, the height at the center increases after a dimensionless waiting period $B$, leading to the liquid lift-off. Thus, $B$ ranges from $0$ to $\infty$. For the case of sessile droplet, however, the edge do not merge, resulting in $H_\mathrm{min}$ becoming the equilibrium thickness $H_\mathrm{eqm}$. The considered slippery system is Krytox, silicone oil, and glass as liquid, lubricant, and solid, respectively, with parameters $r_\mathrm{c}=248\,$\SI{}{\micro\metre}, $\theta=46^\circ$, $h_\mathrm{0}= 6\,$\SI{}{\micro\metre}, and      $\Delta P_\mathrm{liq}=40$ and 95 N/m$^2$ for the liquid bridge and the droplet, respectively.
}
    \label{numerical_spatial_temporal}
\end{figure}

\clearpage
To unveil the lubricant dynamics, we numerically solved the thin-film equation (Methods) \cite{slattery1992,nguyen2003colloidal,Leal2007,softmatterreview_draining}
\begin{equation}
	\frac{\partial h}{\partial t}=\frac{1}{3 r \mu} \frac{\partial}{\partial r}\left(r h^3 \frac{\partial P_{\text {total}}}{\partial r}\right)
	\label{reynold}
\end{equation}
where $\mu$ is the viscosity of oil and $P_{\text {total}}$ is the total pressure in the oil film.  We denote the dimensionless forms of $h, r$, and $t$ as $H(=h/h_0)$, $R(=r/r_\mathrm{c})$, and $T(=t/t_\mathrm{c})$, respectively, where $t_\mathrm{c}=3\mu r_\mathrm{c}^4/\gamma_\mathrm{OA} h_0^3$ is the characteristic time scale  \cite{vella}. From the numerical results, we found that lubricant dynamics in the early-intermediate times (defined as the interval in which most experiments are conducted), proceed through three stages (Fig. \ref{numerical_spatial_temporal}).

%\section{Results}\label{sec2}

%\subsection{Lubricant dynamics in stage-1}
\textbf{Dynamics in stage-I.} In the initial stage, I, the formation of large negative Laplace pressure wetting ridge around $r=r_\mathrm{c}$ causes the oil to rapidly flow from the surrounding area toward the wetting ridge, resulting in a fast decrease in the height at the endpoints of the wetting ridge. The edge height falls faster than the neck height, as the oil beneath the liquid is driven by an extra pressure exerted by the liquid. The rapid decrease causes a non-uniform flow rate, which results in undulations beneath and beyond the liquid. The beneath undulations quickly smooth out to form a wimple profile and the oil continues to flows toward the wetting ridge from both beneath and beyond the liquid. As time progresses, the maxima in the wimple propagate toward the center resulting in a transition from wimple to dimple. This results in the rise of oil thickness at the center with the maximum thickness $h_\mathrm{max,1}$ even exceeding $h_0$, which is also consistent with our experimental observation for the case of $h_{0,1}$. Numerically, we found that $h_\mathrm{max,1}$ depends upon the oil volume present beneath the liquid during merging of the maxima (Supplementary Section 9).

\textbf{Dynamics in stage-II.} After the center reaches a maximum thickness, the dynamics enter stage-II, where the height at the center start decreasing. The result is similar to the film drainage when a bubble or a liquid is pushed against a flat wall in another ambient liquid medium  \cite{softmatterreview_draining,manica_pccp}.
During this period, the wetting ridge continues to grow, with the positions of the edge and the neck propagating toward and away from the center, respectively, while the height at the center and at the edge continues to decrease.  Since the global minimum of the film is at the edge, it is the edge that reaches the van der Waals range ($h\sim100$ nm) the quickest. Here, the stability of the film, and hence the minimum value $h_\mathrm{min}$, depends upon the sign of the disjoining pressure. If the disjoining pressure is repulsive, it balances the Laplace pressure of the liquid, leading to  $h_\mathrm{min}\sim(A_\mathrm{LOS}\,r_\mathrm{c}/\gamma_\mathrm{LA})^{1/3}$ and the oil beneath liquid continues to drain until the thickness at the center reaches the thickness at the edge \cite{oleoplaning,vella}. Here, $A_\mathrm{LOS}$ is the Hamaker constant representing the interaction of liquid and solid phases across the oil phase. Substituting typical values, $h_\mathrm{min}$ calculates to the order of tens of nanometers.  This explains the experimental observation of the minimum height of about 22 nm located at the edge for the case of $h_{0,2}$ and also aligns with the previous study \cite{oleoplaning}. If the disjoining pressure is, however, attractive, the oil film ruptures at the edge and results in $h_\mathrm{min}\sim0$ nm. Thus, for systems with unstable thin-films beneath sessile liquids,
the hole rupture should occur at the edge, which we also observed experimentally on non-modified glass substrates (Supplementary Fig. 4).

\textbf{Dynamics in stage-III.} Subsequent to stage-II, the film dynamics for slippery systems progress to the final stage, stage-III. In this stage, the lubricant dynamics depend upon the equilibrium width of the oil-liquid interface and differ significantly for liquids with different Laplace pressures. For the case of large Laplace pressure liquid bridges and sessile droplets, the inward propagation of the edge stops and leads to the minimum thickness becoming the equilibrium thickness, i.e., $h_\mathrm{min}=h_\mathrm{eqm}$. Conversely, for the case of low Laplace pressure liquid bridges, the edge continue to propagate inward and eventually merge after a time period $b$.
The merging of edge results in the formation of a concave curvature of the oil-liquid interface (as seen from the oil phase), creating a pressure difference that drives the oil back toward the center. This leads to the lift-off of liquid from the slippery surface, explaining the experimental observation for the case of $h_{0,2}$. As the height at the center continues to rise, the system gradually approaches equilibrium. Interestingly, this phenomenon resembles the capillary rise of liquids in thin capillary tubes \cite{jurin_1718,capillarity_2004}.

\begin{figure}[h]
    \centering
    \includegraphics[width=8.5cm]{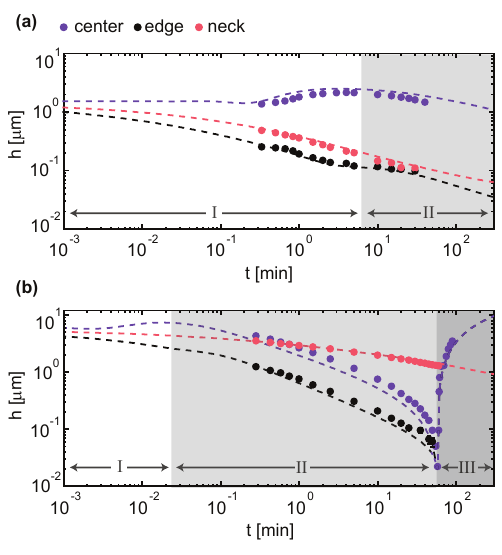}
    \caption{\textbf{Temporal evolution of lubricant film beneath sessile liquid bridge.} Variation of film height $h$ at features: center, edge, and neck with time $t$ beneath sessile Krytox on a silicone oil coated glass with initial film thickness $h_{0,1}=1.5\,$\SI{}{\micro\metre} and $h_{0,2}=5.9\,$\SI{}{\micro\metre}. The first two stages of the dynamics are observed for $h_{0,1}$, while the last two stages are observed for $h_{0,2}$. It takes about 4 min for the center to reach a maximum value of about $2.1$\SI{}{\,\micro\metre} for the case of $h_{0,1}$ and about 58 min to reach a minimum value of about $h_\mathrm{min}= 22$ nm for the case of $h_{0,2}$. The experimental observations for the features are represented by filled circles and the corresponding numerical solutions are shown by dashed lines, with same color as the feature.
}
    \label{temporal_exp}
\end{figure}

The dashed lines in  Fig. \ref{spatial_exp} and  Fig. \ref{temporal_exp} show the numerical solutions for the experimental cases $h_{0,1}$ and $h_{0,2}$ (see Methods for details). For the case of $h_{0,1}$, numerical solutions predict the transition from stage-I to II at about $4$ min and from stage-II to III at about $10^4$ min.  For the case of $h_{0,2}$, the corresponding transitions are predicted at $0.02$ min and $58$ min, respectively. Thus, the time scales involved in stages-I and III are significantly different and make it experimentally challenging to capture all the three stages for a single system within a reasonable time frame. This is why we could not observe the height at the center reaching the van der Waals range (a feature of stage-III) for the case of $h_{0,1}$ and the formation of a wimple in the thin-film (a feature of stage-I) for the case of $h_{0,2}$. The two thicknesses allowed us to partially access the three stages, with the thinner one allowing us to access the first two stages and the thicker one allowing us to access the last two stages. Another alternative would be to vary the width of three-phase contact line, while keeping the initial thickness fixed. This also indicates that the lubricant dynamics and consequently the lubricant depletion can be suppressed by employing slippery systems with thinner films and foreign liquid droplets with larger volumes (Supplementary Section 12).

\clearpage
\section{Dynamics beyond sessile liquids}
\begin{figure}[h]
	\centering
 	\includegraphics[width=8.5cm]{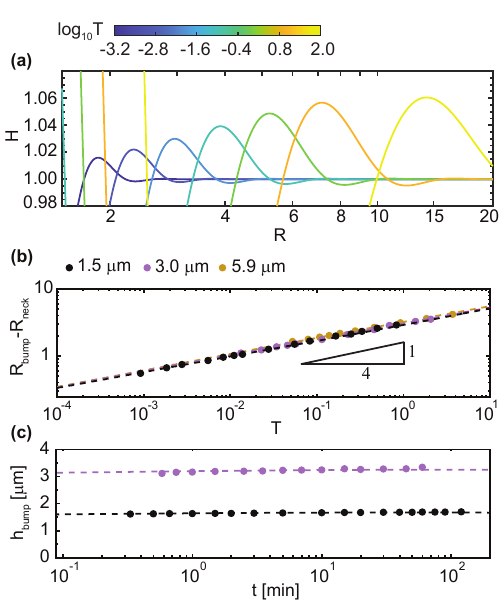}
	\caption{\textbf{Evolution of lubricant film close to bump.} (a) Numerical solution for the evolution of the dimensionless film profile close to the bump. The colorbar corresponds to the dimensionless time $T$. (b) Change in the dimensionless position of the bump relative to the neck $R_\mathrm{bump}-R_\mathrm{neck}$ as a function of $T$ for $h_0= 1.5, 3.0$, and $5.9\SI{}{\,\micro\metre}$. The rescaled bump position follows power-law, $R_{\mathrm{bump}}-R_{\mathrm{neck}} \sim T ^{1/4}$. (c)  Variation in height at the bump $h_\mathrm{bump}$ with time $t$ for $h_0=$ $1.5$ and $3.0\SI{}{\,\micro\metre}$ (see Supplementary Section 10 for details on individual features).  For both (b) and (c), the filled circles correspond to the experimental observations, while the dashed lines (same color as the corresponding experiment) represent numerical solution. 
	}
	\label{bump}
\end{figure}

Along with the thin-film dynamics beneath sessile liquids, we also analyzed the dynamics of thin-film close to the bump. We found that throughout the three stages, the bump propagates outward according to the scaling law,  $R_\mathrm{bump}-R_\mathrm{neck} \sim T ^{1/4}$ without significant variation in the film height at the bump $h_\mathrm{bump}$. Here, $R_\mathrm{bump}$ and $R_\mathrm{neck}$ are the dimensionless positions of the bump and neck, respectively. A similar scaling law has also been  reported previously for capillary driven flow of thin viscous films \cite{capillary_healing_2018,vella,drawingliquidbridge}. Fig. \ref{bump}(b),(c) illustrates the observed behavior found from experiments and numerical solutions (dashed lines) for the cases $h_{0,1}$ and $h_{0,2}$ discussed previously, along with a new case $h_{0,3}$. 
For the case $h_{0,3}$, the system parameters were $\Delta P_\mathrm{liq} \sim$ 42\,$\SI{}{N/m^2}, r_\mathrm{c}\sim 153\,\SI{} {\micro\metre}$, and initial film thickness $h_{0,3}=3.0\,\SI{} {\micro\metre}$. We also numerically found same scaling law for the case of sessile droplets, demonstrating the universality of the scaling law for sessile liquids.

To derive the scaling law, we refer to the approach adopted by Gonzalez \textit{et al.} \cite{drawingliquidbridge}. First, we consider the dominant terms contributing in the region close to the bump in equation (\ref{reynold}) to obtain 
\begin{equation}
	\frac{\partial h}{\partial t}=\frac{-\gamma_{\mathrm{OA}}}{3 r \mu} \frac{\partial}{\partial r}\left(r h^3 \frac{\partial}{\partial r} \left(h_{r r}+\frac{h_r}{r}\right)\right).
\end{equation}
Next, since the bump propagates with nearly a constant height, we write $h=h_\mathrm{bump}+\epsilon$ where $\epsilon \ll h_\mathrm{bump}$. Finally, we approximate the scaling of $\nabla\sim 1/( r_\mathrm{bump}-r_{\mathrm{neck}})$ and substitute the physical quantities in their dimensionless form to obtain the observed scaling law.

\section {Configuration at equilibrium}
\begin{figure}[h]
	\centering
	\includegraphics[width=8.5cm]{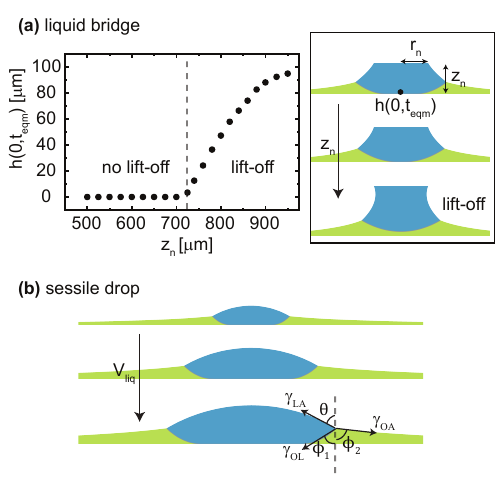}
	\caption{\textbf{Equilibrium configuration of system.} (a) Variation of equilibrium height at the center $h(0,t_\mathrm{eqm})$ as a function of needle height $z_\mathrm{n}$ with fixed needle radius $r_\mathrm{n}=$ \SI{550}{\micro\metre} for Krytox volume $V_\mathrm{liq}$ = \SI{1}{\micro\litre}. The liquid lift-off occurs for $z_\mathrm{n}> $ \SI{ 700}{\micro\metre}. The equilibrium configurations for $z_\mathrm{n}=600, 700, \textrm{and } 1000$ \SI{}{\,\micro\metre}	arranged from top to bottom are displayed in the right panel. A similar behavior for lift-off is also observed with variation in $r_\mathrm{n}$ (Supplementary Fig. 10). (b) Equilibrium configurations of the system upon contact with a sessile Krytox droplet with $V_\mathrm{liq}$ = 10, 50, and 100 \SI{}{\nano\litre} (arranged from top to bottom) on a silicone oil coated glass. For the case of a liquid bridge, the edge merge at low liquid pressure, whereas for a sessile droplet, the edge separate further as the pressure decreases.  $\theta, \phi_1$, and $\phi_2$ are the angles at the three phase contact line that the interfacial tensions $\gamma_\mathrm{LA}$, $\gamma_\mathrm{OL}$, and $\gamma_\mathrm{OA}$, respectively, make with the vertical. The configurations are numerically computed using Surface Evolver software (see Supplementary Section 11 for details). 
	}
	\label{SE_eqm}
\end{figure}

To investigate the effect of Laplace pressure of liquid on the equilibrium configuration, we employed Surface Evolver (Methods and Supplementary Section 11 for more details) \cite{SE, surfaceevolvermanual,semprebon_2017,shivam_2023}. We varied the needle height relative to the surface $z_\mathrm{n}$ for the case of liquid bridges and the volume of liquid $V_\mathrm{liq}$ for the case of droplets. We found results similar to those predicted by our theoretical model for thin-film dynamics. For the case of a liquid bridge, we found that the merging of edge occurs for liquids with very low Laplace pressure. Fig. \ref{SE_eqm}(a) illustrates one such example for Krytox, silicone oil, and glass system.  With fixed values of $V_\mathrm{liq}$ = \SI{1}{\micro\litre} and  $r_\mathrm{n}=$ \SI{550}{\micro\metre}, lift-off occurred when $\Delta P_\mathrm{liq}$ was reduced down to $13$ N/m$^2$ for $z_\mathrm{n}> $ \SI{ 700}{\micro\metre}. A similar behavior was observed upon varying  $r_\mathrm{n}$ with fixed  $z_\mathrm{n}$ (Supplementary Fig. 10).  For the case of a sessile droplet, however, we did not observe the merging of edge at any value of Laplace pressure. Instead, the distance from edge to the center increased with a decrease in Laplace pressure (Fig.\ref{SE_eqm}(b)).

Analytically, the non-occurrence of lift-off for the sessile droplets can be understood by considering the possibility of merging of the edge. Let $\phi_1$ and $\phi_2$ be the angles that the interfacial tensions $\gamma_\mathrm{OL}$ and $\gamma_\mathrm{OA}$, respectively, make with the vertical. If the edge were to merge, the oil-droplet interface would follow a spherical cap profile (neglecting hydrostatic pressure) with a Laplace pressure of $2\gamma_\mathrm{OL} \cos{\phi_1}/r_\mathrm{c}$. With $P_\mathrm{oil}=P_\mathrm{atm}$, the Laplace pressure of the droplet-air interface, $2\gamma_\mathrm{LA} \cos{\theta}/r_\mathrm{c}$, would need to be equal to $2\gamma_\mathrm{OL} \cos{\phi_1}/r_\mathrm{c}$. The balance of vertical forces at the three-phase contact line, $\gamma_\mathrm{LA} \cos{\theta}=\gamma_\mathrm{OL} \cos{\phi_1}+\gamma_\mathrm{OA} \cos{\phi_2}$ would then require $\phi_2=90^\circ$, which is, in general, implausible. Therefore, for the case of sessile liquid droplets, the edge do not merge, and droplet lift-off does not occur. This is in contrast for liquid bridges, for which there is no such restriction on Laplace pressure and therefore, lift-off occurs once $\Delta P_\mathrm{liq}\sim2\gamma_\mathrm{OL} \cos{\phi_1}/r_\mathrm{c}$.

\clearpage
\section{Conclusions}
Using an unconventional slippery system of Krytox, silicone oil, and glass, we have provided a detailed understanding of the thin-film lubricant dynamics upon contact with sessile liquids on slippery surfaces. 
The developed understanding of lubricant dynamics beneath sessile liquids enables the prediction of heat flow in condensation-based heat exchangers, while that beyond liquids facilitates control over droplet coalescence.
The developed theoretical model captures fine details of system dynamics and can be adapted to explore other aspects, including loss of lubricant in the form of wetting ridge, which is one of the major limitations of slippery surfaces.  
The methods employed in this study may also provide a framework for investigating the dynamics of thin-films on lubricant infused slippery surfaces, which are commercially employed versions of slippery surfaces.  Altogether, the study reveals interfacial behaviors that were previously inaccessible, offering a deeper perspective on the mechanisms governing lubricant dynamics.

\section{Acknowledgements}
We thank V. J. Yallapragada for fruitful discussions. The project has received funding support from Department of Science and Technology, New Delhi (Project No. DST/PHY/2022523) under the DST-FIST program. 
S.G. acknowledges financial support from the Prime Minister's Research Fellows (PMRF) scheme of the Government of India.

\section{Data and code availability}
The data and code supporting the plots and findings presented in this study are available from the corresponding author upon reasonable request.

\clearpage
\section{Methods}\label{sec2}

\textbf{Sample preparation.} We used glass coverslips as the solid substrates. Silicone oil (Gelest) with kinematic viscosity $1000\times10^{-6}\, $\SI{ }{m^2/s}, density 971 Kg/m$^3$ and surface tension 21 mN/m, was used as the lubricant.For the non-evaporating liquid with a high refractive index contrast with silicone oil, we chose Krytox VPF-1506 (Chemours) with kinematic viscosity $60\times10^{-6}\, $\SI{ }{ m^2/s}, density 1880 Kg/m$^3$, and surface tension $17.0\pm0.5$ mN/m as the liquid.  The interfacial tension value for the oil-liquid interface ($\gamma_{\mathrm{OL}}=$ $8.0\pm0.8$ mN/m) was determined using the pendant drop method, conducted with a contact angle goniometer (DataPhysics OCA 35) \cite{pendant_drop_method}. 

For the system of Krytox, silicone oil, and glass, while $A_\mathrm{LOS}>0$, the low contact angle of Krytox on glass results in $S_\mathrm{OSL}(=\gamma_{\mathrm{LS}}-\gamma_{\mathrm{OL}}-\gamma_{\mathrm{OS}}) < 0$ (Supplementary Section 4). However, for a thin-film of lubricant to be stable, the system must satisfy  $A_\mathrm{LOS}>0$ and $S_\mathrm{OSL} > 0$ \cite{oleoplaning}. Thus, to decrease the wettability of Krytox on glass, the substrates were  chemically modified by grafting polydimethylsiloxane (PDMS) chains. The procedure is as follows \cite{universal,mccarthy_2010,mccarthy_2011}. First, the substrates were exposed to oxygen plasma (Harrick Plasma) for 15 min, then spin coated with silicone oil, and subsequently heated at 150$^\circ$C for 10 h. Afterwards, the non-grafted oil was removed by first rinsing and then ultrasonicating the samples in toluene (99.5 $\%$, Loba Chemie) for 10 min.

\textbf{Determining thickness of grafted  chains.} We employed X-Ray Reflectivity (XRR) technique (PANanalytical X'pert Pro, $\lambda=1.54~$\AA) to determine the thickness of the grafted PDMS chains. By analyzing the positions of the maxima and minima in the reflectivity data, we determined the approximate thickness of the chains to be about 7 nm.  The details can be found in the Supplementary Section 5. 

\textbf{Determining initial film thickness.}
Glass coverslips with a few \SI{}{\micro\liter} oil droplets dispensed on top, were spun at varying angular speeds to achieve different film thicknesses using a commercial spin coater (spinXG-P1, Apex). Specifically, the coverslips were spun at 4300 rpm for 100 s, 8100 rpm for 100 s, and 10,000 rpm for 240 s to achieve thicknesses of $5.9$, $3.0$, and $1.5\SI{}{\,\micro\metre}$, respectively, with an uncertainty of $\pm0.1\SI{}{\,\micro\metre}$. The thicknesses were measured using an in-house developed thickness measurement setup employing an inverted microscope (IX-73, Olympus), a spectrometer (HR-4000, Ocean Optics) and a metal halide light source (U-HGLGPS, Olympus). A 4X magnification objective with a 0.1 numerical aperture (NA) was used to focus the light on the sample. See Supplementary Section 1 for more details. 

\textbf{Side-view imaging.} A halogen light source (Radical Scientific Equipments Pvt. Ltd.) and a zoom lens (Edmund Optics) were employed to observe the system from the side. The default height of the confocal sample stage, however, does not permit a side view. To address this, we designed a custom acrylic stage with a height of about 8 mm and a through hole of about 15 mm diameter. This elevated the sample, allowing the system to be viewed simultaneously from both the side and bottom.

\textbf{Bottom-view imaging.} 
The thin-film interference was observed using a confocal microscope (Leica Stellaris 5) in the reflection mode. Blue and red lasers with wavelengths $\lambda_{1}=488 \text{ nm}$ and $\lambda_{2}=638 \text{ nm}$, respectively, were simultaneously incident at the sample, and the reflected light corresponding to the two lasers was collected by two separate detectors. The light was focused on the thin-film from a 10X objective with 0.3 NA. The pinhole size was set to 2 airy units. The image was captured at a scan speed of 600 Hz with the pixel dwell time of about 0.425 µs. The resolution of the image was 2048 x 2048 pixel with a pixel size of 0.757 µm. The physical size of the image was 1550 µm along both the horizontal and vertical directions. The laser intensities were adjusted such that both the inner ($r<r_\mathrm{c}$) and outer interference fringes ($r>r_\mathrm{c}$) were clearly visible simultaneously.

\textbf{Determining Laplace pressure of liquid.} 
The Laplace pressure of the liquid was determined by employing Surface Evolver,  a finite element software that computes the equilibrium liquid configuration subject to various constraints  \cite{SE,surfaceevolvermanual}. From the known constraints and values of $z_\mathrm{n}$, $r_\mathrm{n}$, and $r_\mathrm{c}$, the configuration and hence the Laplace pressure of the liquid bridge were calculated. The details can be found in the Supplementary Section 6. 

\textbf{Determining dynamic film thickness.} 
To determine the thin-film profile, we employed the dual wavelength reflection interference contrast microscopy (DW-RICM) technique \cite{dwricm_1964,sengupta_1,sengupta_2,oleoplaning}. Briefly, a pair of normalized light intensities at a region of interest was overlapped on the theoretical Lissajous  curve and the corresponding thickness was determined (see Supplementary Section 3 for details).

\textbf{Solving thin-film dynamics.}  
We employed a two-dimensional axisymmetric model in the $r-z$ plane
to theoretically analyze the dynamics of thin-films both beneath and beyond the liquid. For slippery systems, the length scale along $r$ is generally much larger than that along $z$, and the no-slip boundary condition is satisfied at the oil-solid interface. Thus, we employed the Stokes-Reynolds equation in cylindrical coordinates. The equation relates the rate of change of thickness $\partial h(r,t)/\partial t$ to $P_{\text {total}}$ and $\mu$ of the oil film as  \cite{slattery1992,nguyen2003colloidal,Leal2007,softmatterreview_draining}
 \begin{equation}
	\frac{\partial h}{\partial t}=\frac{\beta}{12 r \mu} \frac{\partial}{\partial r}\left(r h^3 \frac{\partial P_{\text {total}}}{\partial r}\right)
	\label{reynold2}
\end{equation} 
where the value of constant $\beta$ depends upon the boundary condition at the oil-liquid interface. The value of $\beta=4$, if a zero tangential stress boundary condition is satisfied at the oil-liquid interface, while $\beta=1$ if no-slip or tangentially immobile boundary condition is satisfied at the interface.
For the present case, the viscosity of liquid is about ten times smaller than that of the lubricant. Thus, we considered $\beta=4$.
Due to the presence of liquid in the region, $r\le r_\mathrm{c}$, the total pressure within the oil film, relative to atmospheric pressure, differs across the regions $r<r_\mathrm{c}$, $r=r_\mathrm{c}$, and $r>r_\mathrm{c}$ and can be expressed as
\begin{equation}
	\begin{aligned}
		P_\mathrm{total}=
		&\begin{aligned}
			& \left[\Delta P_{\mathrm{liq}}-\gamma_{\mathrm{OL}}\left(h_{r r}+\frac{h_r}{r}\right)-\frac{A_{\mathrm{LOS}}}{6 \pi h^3}\right] \mathrm{U}\left(r_{\mathrm{c}}-r\right) \\
		\end{aligned}\\
		&\left[-\gamma_{\mathrm{OA}}\left(h_{r r}+\frac{h_r}{r}\right)-\frac{A_{\mathrm{AOS}}}{6 \pi h^3}\right]  \mathrm{U}\left(r-r_{\mathrm{c}}\right)
		-\gamma_{\mathrm{LA}} \cos \theta ~\delta\left(r-r_{\mathrm{c}}\right)\\
		&+ \mathrm{\rho g }(h-z)
	\end{aligned}
	\label{eqn_pressure}
\end{equation}
\noindent where $A_{\mathrm{AOS}}$ is the non-retarded Hamaker constant corresponding to the interaction between the air and solid phase across the oil phase. For the sessile droplets, $\Delta P_{\mathrm{liq}}=2 \gamma_\mathrm{LA} \cos\theta/\mathrm{r_{c}}$, while for the liquid bridges, it needs to be determined from the side view. The coefficient $U$ is the unit step function and $\theta$ is the angle that  $\gamma_{\mathrm{LA}}$ makes with the vertical. We assumed small slopes in the oil film, $\lvert h_r \rvert \ll 1 $ to approximate the mean curvature of the oil-liquid and oil-air interfaces. To numerically solve the Stokes-Reynolds equation, 
we non-dimensionalized $h, r$, and $t$ using the characteristic scales for the early-intermediate times $h_0$, $r_\mathrm{c}$, and $t_\mathrm{c}$, respectively, and applied the following boundary conditions for $h(r,t)$ \cite{vella}:
\begin{equation}
	h_r(0,t)=h_r(r_{\infty},t)=h_{rrr}(0,t)=h_{rrr}(r_{\infty},t)=0
\end{equation}
where $r_{\infty}$ is the position far from the liquid. The scales for early-intermediate time were particularly chosen because this time encompasses the conditions of most experimental observations.  Using the central finite difference technique and method of lines, a set of ODEs was obtained for equation (\ref{reynold2}), which was solved numerically using MATLAB’s built-in solver ode15s. The connection of the inner and the outer menisci was achieved by using the smoothed form of delta and step functions similar to the method described by Zhaohe \textit{et al.} \cite{vella}. The details can be found in the Supplementary Section 8. 

\textbf{Fitting experimental results with numerical predictions.} 
To fit the experimental observations with numerical results, we considered the average values of $r_{\mathrm{c}}$ and $\Delta  P_\mathrm{liq}$ and varied the values of $\gamma_{\mathrm{OL}}$ within the uncertainty limits ($\gamma_{\mathrm{OL}}=8.0\pm0.8$ mN/m) and $\theta$ within the dynamic range 
($\theta=44^\circ-47^\circ$). We found good quantitative agreement for $\gamma_{\mathrm{OL}}=$ 8.7 mN/m and $\theta=46^\circ$ and $47^\circ$ for $h_0= 5.9\,\SI{}{\micro\metre}$ and $ 1.5\,\SI{}{\micro\metre}$, respectively. A more refined approach would be to dynamically update $r_\mathrm{c},\Delta P_\mathrm{liq},$ and $\theta$ over time, an aspect that future studies may address.

\textbf{Determining equilibrium configuration.} 
The equilibrium configuration of the system was determined by employing Surface Evolver. We set the Laplace pressure of oil at $z=0$ equal to zero and let the system evolve with the upper liquid bridge vertices held onto the circular boundary of fixed radius $r_\mathrm{n}$ (Supplementary Section 11). We did not consider intermolecular interactions, as they are expected to primarily govern the stability of thin-films and the time required to reach equilibrium, rather than the equilibrium macroscopic profile  \cite{vella}.

\clearpage
\bibliography{references}% common bib file

\end{document}

\author{
	Shivam Gupta\textsuperscript{1}, 
	Bidisha Bhatt\textsuperscript{1}, 
	Zhaohe Dai\textsuperscript{2,*}, and
	Krishnacharya Khare\textsuperscript{1,*} \vspace{1em}
}
%\email{kcharya@iitk.ac.in}

\affiliation{\textsuperscript{1}Department of Physics, Indian Institute of Technology Kanpur, Kanpur 208016, Uttar Pradesh, India \\ \textsuperscript{2}Department of Mechanics and Engineering Science, State Key Laboratory for Turbulence  and \\ Complex Systems, College of Engineering, Peking University, Beijing 100871, China}
%\email{kcharya@iitk.ac.in}
%\title[Article Title]{Dynamics of Thin Lubricant Films upon Liquid Contact on Slippery Surfaces}

\title{
	Supplementary information for\\[3em]
	%\rule{\textwidth}{0.4pt} \\
	%\rule{\textwidth}{0.4pt} \\[1em]
	\textbf{\Large Dynamics of Thin Lubricant Films upon Liquid Contact on Slippery Surfaces}
}

\maketitle
{
	\setlength{\baselineskip}{2pt} % Adjust: try 10pt, 11pt, or 12pt for tighter spacing
	\vfill
	\begin{flushleft}
		\rule{\textwidth}{0.8pt}  % Thick horizontal line (full width, 0.8pt height)
		\small\textit{email: \href{mailto:daizh@pku.edu.cn, kcharya@iitk.ac.in}{daizh@pku.edu.cn, kcharya@iitk.ac.in 
		}}
		
	\end{flushleft}
	
}
%\vspace{5em} % Space you want to create between authors and affiliations
\newpage
\section*{Contents}
\begin{itemize}
	\item[\textbf{S1.}] Determining the initial film thickness
	\item[\textbf{S2.}] Setup for simultaneous side and bottom observation
	\item[\textbf{S3.}] Determining the dynamic film thickness
	\item[\textbf{S4.}] Calculating the Hamaker constant and the spreading coefficient
	\item[\textbf{S5.}] Determining the thickness of grafted polydimethylsiloxane chains
	\item[\textbf{S6.}] Calculating the Laplace pressure of liquid bridges
	\item[\textbf{S7.}] Variation in the Laplace pressure of liquid and width of contact line
	\item[\textbf{S8.}] Numerical scheme for studying the lubricant dynamics
	\item[\textbf{S9.}] Variation in maximum height at the center in stage-I and waiting period in stage-III
	\item[\textbf{S10.}] Variation in the dimensionless position and height over time
	\item[\textbf{S11.}] Determining the equilibrium configuration
	\item[\textbf{S12.}] Effect of initial film thickness and number of droplets on the size of wetting ridge
	\item[\textbf{}] References
\end{itemize}

\newpage

\clearpage
\section{Determining the initial film thickness}
\begin{figure}[h]
	\centering
	\includegraphics[width=8.5cm]{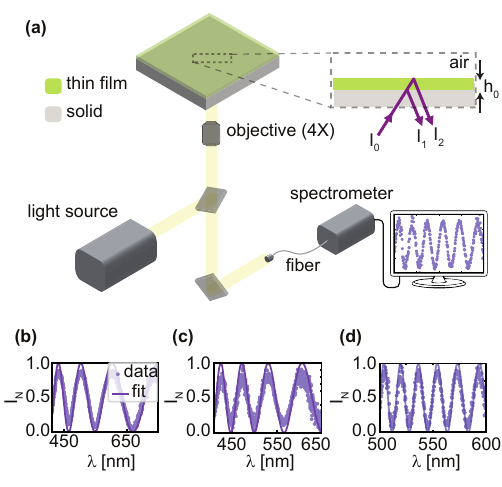}
	\caption{Thickness measurement of thin films. (a) Schematic of the setup used to measure thickness of a thin supported film using an inverted microscope and a spectrometer. The thickness of the film $h_0$ is determined from the intensity information at different wavelengths using equation (\ref{norm_int}). (b) Normalized reflected intensity as a function of wavelength $\lambda$ for a reference sample consisting of a 1.0\SI{}{\,\micro\metre} $\mathrm{SiO_2}$ thermal oxide layer on Si. (c) and (d) show the normalized intensity for a $1.5 \SI{}{\,\micro\metre}$ and a $5.9 \SI{}{\,\micro\metre}$ thin silicone oil film coated on glass coverslips, respectively. In (b)-(d), the filled circles represent experimental data points, while the solid lines correspond to the theory (equation (\ref{norm_int})).
	}
	\label{thick meas}
\end{figure}

We developed an in-house thickness measurement setup to measure the initial thickness of a thin lubricant film $h_0$. The setup consisted of a spectrometer and the built-in optics of the inverted microscope as illustrated in Fig. \ref{thick meas}. Light after passing through the semi-reflective mirror was focused by the objective lens (4X, 0.1 NA) onto the sample and the reflected light was collected by the spectrometer using an optical fiber.  
The reflected light intensity $I_1(\lambda)$ and $I_2(\lambda)$ from the bottom and top interfaces of the thin film, respectively, interfere to give the resultant light intensity $I_r(\lambda)$ given as \cite{born2013principles}
\begin{equation}
	I_r(\lambda)=I_1(\lambda) +I_2(\lambda)+2\sqrt{I_1(\lambda) I_2(\lambda)} \cos{\phi}
	\label{resultant_intensity_eqn}
\end{equation}
where the phase difference $\phi$ between the light intensities is given as
\begin{equation}
	\phi=\begin{cases}4 \pi n_{\mathrm{oil}}\, h_0/ \lambda & n_{\mathrm{substrate}}>n_{\mathrm{oil}}>n_{\mathrm{air}} \textrm{ or }  n_{\mathrm{substrate}}< n_{\mathrm{oil}} < n_{\mathrm{air}}\\\pi+ 4 \pi n_{\mathrm{oil}}\, h_0/ \lambda & \textrm{otherwise}\end{cases}
\end{equation}

\noindent Assuming negligible variation of refractive indices with $\lambda$, $I_1(\lambda)$ and $I_2(\lambda)$ can be written as $I_1(\lambda)= I_0(\lambda) I'_1$ and $I_2(\lambda)= I_0(\lambda) I'_2$, where $I_0(\lambda)$ is the incident light intensity and $I'_1$ and $I'_2$ consists of the Fresnel's reflection and the transmission coefficients. Typically, light sources emit light of different intensities at different wavelengths; thus, $I_0$ is considered a function of $\lambda$. Substituting the expressions for $I_1(\lambda)$ and $I_2(\lambda)$ in equation  (\ref{resultant_intensity_eqn}) gives, 	$I_r(\lambda)=I_0(\lambda)\left(I'_1 +I'_2+2\sqrt{I'_1 I'_2} \cos{\phi}\right)$. The expression can be further simplified by using the reference intensity $I_\mathrm{ref}(\lambda)$, which is the reflected intensity collected from a non-absorbing sample, e.g., glass or silicon. Noting, $I_\mathrm{0}(\lambda)/I_\mathrm{ref}(\lambda)=C$ where $C$ is a constant, the ratio  $I_r(\lambda)/I_\mathrm{ref}(\lambda)=I_r'(\lambda)=C\left(I'_1 +I'_2+2\sqrt{I'_1 I'_2} \cos{\phi}\right)$. The expression can be further simplified by normalizing $I_r'(\lambda)$ by using the maximum ($I_\mathrm{max}$) and the minimum ($I_\mathrm{min}$) intensity
	\begin{equation}
		I_\mathrm{N}=\frac{I'_\mathrm{r}-I_\mathrm{min}}{I_\mathrm{max}-I_\mathrm{min}}=\frac{1+\cos \phi}{2}
		\label{norm_int}
	\end{equation} 
On fitting the experimental data with equation (\ref{norm_int}), the thickness $h_0$ can be determined.
To confirm the accuracy of the setup, we calculated the thickness of a \SI{1.0}{\micro\metre} thermal oxide layer on Silicon (University Wafers Inc.) by placing the opaque sample in an inverted position (Fig. \ref{thick meas}(b)). The measured value was also found to be $1.0 \SI{}{\,\micro\metre}$, confirming the accuracy of the setup. Similarly, the thickness of the thin films spin coated at different rotational speeds can be determined by employing the developed setup as shown in Fig. \ref{thick meas}(b,c). Note that the accuracy will decrease for systems whose refractive indices get significantly affected in the visible light range and/or if the systems are absorptive in nature. Also, the numerical aperture effects have been neglected when calculating the thickness of thin films.
	
%, which becomes significant if large NA objectives are used. 

\clearpage
\section{Setup for simultaneous side and bottom observation}

\begin{figure}[h]
	\centering
	\includegraphics[width=11.4cm]{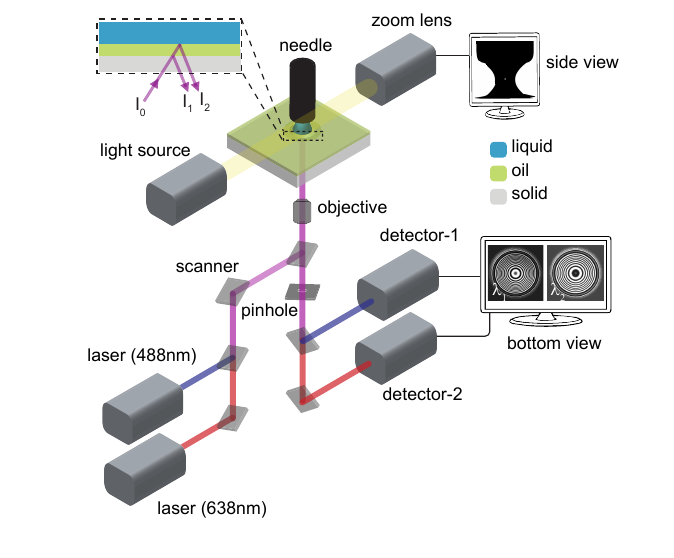}
	\caption{(a) Schematic representation of the setup used to observe lubricant dynamics. The liquid is held stationary by a needle and the system is monitored simultaneously from both side and bottom views. The side view is captured using a light source and a zoom lens (with camera), while a confocal microscope in reflection mode is used for the bottom-view. Two lasers ($\lambda_{1}=488 \text{ nm}$ and $\lambda_{2}=638 \text{ nm}$) are simultaneously directed onto the film, and the reflected light corresponding to the two wavelengths, is collected simultaneously by two separate detectors.
	}
	\label{dwricm setup}
\end{figure}
% , capturing light with wavelengths $\lambda_{1}$ and $\lambda_{2}$, respectively. 

\clearpage
\section{Determining the dynamic film thickness}

\begin{figure}[h]
	\centering
	\includegraphics[width=11.4cm]{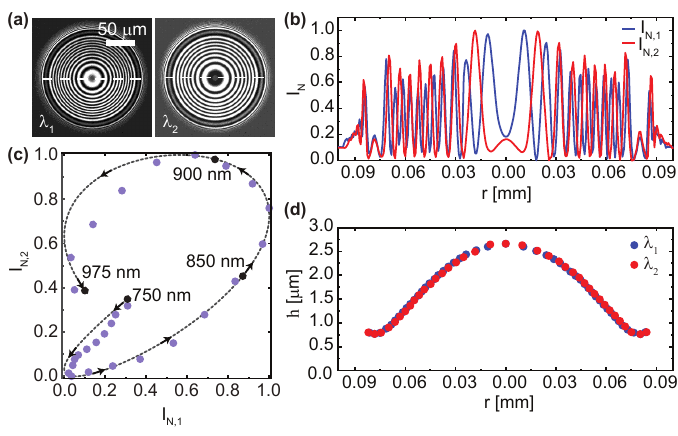}
	\caption{Methodology to determine the lubricant film profile using dual wavelength reflection interference contrast microscopy. First, the interferometry images are simultaneously captured using the two wavelengths $\lambda_{1}$ and $\lambda_{2}$ as shown in (a). Next, a line is drawn across the region of interest (white horizontal dashed lines in (a)) and the normalized intensity profiles are obtained, as shown in (b). The blue and red curves in (b) correspond to the normalized intensity profiles $I_\mathrm{N,1}$ and $I_\mathrm{N,2}$ for the blue ($\lambda_{1}=488 \text{ nm}$) and red laser ($\lambda_{2}=638 \text{ nm}$), respectively. The intensity pairs at the edge are then mapped onto the theoretical Lissajous curve (equation \ref{norm_int2}) to determine the thickness at the edge, as depicted in (c). For the example shown, the experimental normalized intensity pair (filled circles) follows the theoretical Lissajous curve ranging from 750 nm to 975 nm (black dashed curve), implying the thickness at the edge is 750 nm. The black filled circles represent the thickness values from the theory. The thickness is then adjusted by $\lambda/4 n_\mathrm{oil}$ based on the positions of adjacent maxima and minima. Finally, the thickness at the center is determined again using the Lissajous curve (not shown here). Thus, the complete lubricant profile is reconstructed using the two wavelengths, as shown in (d). The blue and red filled circles represent the thickness values determined from the corresponding wavelengths. 
	}
	\label{dwricm_tech}
\end{figure}

To measure the dynamic thickness, we fixed the wavelength of light and collected the reflected light intensity at the region of interest. The resultant reflected light intensity is still given by equation (\ref{resultant_intensity_eqn}) (with $n_\mathrm{liquid}$
in place of $n_\mathrm{air}$ beneath liquid), however, with fixed $\lambda$, a reference intensity is not required and the normalized intensity $I_\mathrm{N}$ can be directly determined using the maximum ($I_\mathrm{max}$) and minimum ($I_\mathrm{min}$) intensities to obtain  \cite{oleoplaning}
\begin{equation}
	I_\mathrm{N}=\frac{1+\cos \phi}{2}
	\label{norm_int2}
\end{equation}
Since $I_\mathrm{N}$ repeats after every $\lambda/2\,n_{\mathrm{oil}}$, determining the thickness using only one wavelength results in ambiguity. The ambiguity can be removed if the reflected light intensity is collected simultaneously using two or more wavelengths. The technique using two wavelengths is called Dual-Wavelength Reflection Interference Contrast Microscopy. From the distinct values of $I_\mathrm{N,1}$ and $I_\mathrm{N,2}$ corresponding to two distinct wavelengths $\lambda_1$ and $\lambda_2$, the thickness $h$ is determined. We used blue and red lasers with wavelengths $\lambda_1=488$ nm and $\lambda_2=638$ nm, respectively, to obtain $I_\mathrm{N,1}$ and $I_\mathrm{N,2}$. The experimental data is then plotted on the theoretical Lissajous curve to determine the value of $h$ for the corresponding intensity pair, as shown in Fig. \ref{dwricm_tech}.

\clearpage

\clearpage

\section{Calculating the Hamaker constant and the spreading coefficient}
The value of the Hamaker constant is calculated using the non-retarded theory, expressed as \cite{israelachvili}
\begin{equation}
	A_\mathrm{LOS} = -\frac{3}{4} k_\mathrm{B} T \left( \frac{\epsilon_\mathrm{O} - \epsilon_\mathrm{L}}{\epsilon_\mathrm{O} + \epsilon_\mathrm{L}} \right) \left( \frac{\epsilon_\mathrm{L} - \epsilon_\mathrm{O}}{\epsilon_\mathrm{L} + \epsilon_\mathrm{O}} \right) + \frac{3 \pi \hbar \nu_\mathrm{e}}{4 \sqrt{2}} 
	\frac{(n_\mathrm{O}^2 - n_\mathrm{L}^2)(n_\mathrm{S}^2 - n_\mathrm{O}^2)}{\sqrt{(n_\mathrm{L}^2 + n_\mathrm{O}^2)(n_\mathrm{S}^2 + n_\mathrm{O}^2)} \left[ \sqrt{n_\mathrm{L}^2 + n_\mathrm{O}^2} + \sqrt{n_\mathrm{S}^2 + n_\mathrm{O}^2} \right]}
	%\tag{1}    
\end{equation}
where $\nu_{e} \approx 4 \times 10^{15} \,\mathrm{s}^{-1}$ is the mean absorption frequency. $k_\mathrm{B}$ and $T$ are the Boltzmann constant and the absolute temperature, respectively. $\epsilon_i \textrm{ and } n_i $ are the dielectric permittivity and refractive index of the $i$-th material, respectively, with $i$  representing oil, liquid, and solid. For the system of Krytox, silicone oil, and glass, using the known values of dielectric permittivities and refractive indices (table \ref{table}), $A_\mathrm{LOS}$ calculates to $4\times 10^{-21}\mathrm{\,J}$. Thus, the system satisfies the first criterion for the stability of a lubricating film $A_\mathrm{LOS}>0 $.

\begin{figure}[h]
	\centering
	\includegraphics[width=8.5cm]{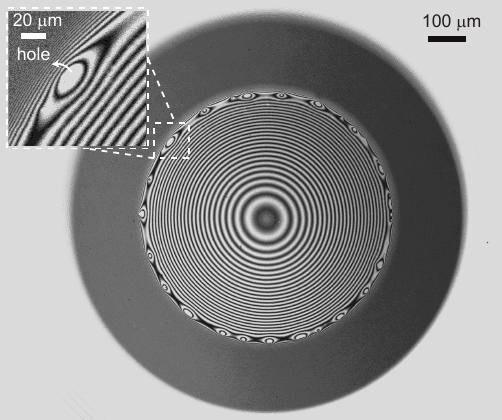}
	\caption{Interferometry image of a silicone oil film, coated on an unmodified glass, upon contact of a sessile Krytox. The image is obtained by using a confocal microscope set in the reflection mode ($\lambda=638 \text{ nm}$). Since the system satisfies $S_\mathrm{OSL}<0,$ Krytox destabilizes the film via hole rupture. Notably, all the holes appear at the edge. The inset shows the zoomed-in image of one of the ruptured holes.
	}
	\label{dewetting}
\end{figure}

\begin{table}[h!]
	\centering
	\caption{Optical and dielectric properties of the materials. The subscripts $\mathrm{S,O},$ and $\mathrm{L}$ indicate silicone oil, glass and Krytox, respectively.}
	\begin{tabular}{c c c c c c c}
		\hline \hline
		$\epsilon_\mathrm{S}$ & $n_\mathrm{S}$ & $\epsilon_\mathrm{O}$ & $n_\mathrm{O}$ & $\epsilon_\mathrm{L}$ & $n_\mathrm{L}$ \\ \hline
		7.0 & 1.51 & 2.6 & 1.41 & 2.1 & 1.3 \\ \hline
	\end{tabular}
	
	\label{table}
\end{table}
\noindent To check the second stability criterion, we evaluated the contact angle of liquid on solid in ambient oil defined as $\theta_\mathrm{LSO}$. Using Young's equation, the expression for $S_\mathrm{OSL}$ simplifies to $S_\mathrm{OSL}=\gamma_\mathrm{OL}(1+\cos{\theta_\mathrm{LSO}})$  \cite{oleoplaning}. Thus, if $\theta_\mathrm{LSO}=180^\circ$, only then the system satisfies the second criteria. For untreated glass coverslips, it was found that  $\theta_\mathrm{LSO}<180^\circ$, resulting in $S_\mathrm{OSL}<0$ and thus Krytox dewets the thin film of oil as shown in Fig. \ref{dewetting}. Note that all the holes emerge at the edge. To increase the angle $\theta_\mathrm{LSO}$, the glass surfaces were grafted with PDMS chains. It was found that for the modified glass-silicone oil-Krytox system, the angle $\theta_\mathrm{LSO}\sim180^\circ$ implying $S_\mathrm{OSL}>0$. This was also confirmed by the presence of stable thin lubricant film throughout the interferometry measurements for such systems.
Thus, the system satisfies both stability criteria, making this combination suitable for studying long time lubricant dynamics.

\clearpage
\section{Determining the thickness of grafted polydimethylsiloxane chains}

\begin{figure}[h]
	\centering
	\includegraphics[width=8.5cm]{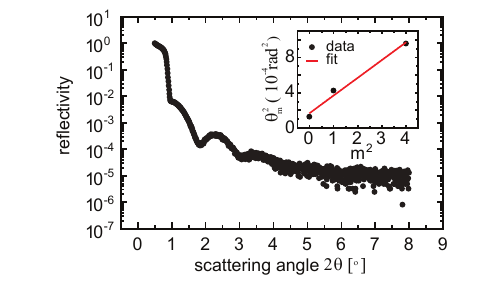}
	\caption{Variation of reflectivity with scattering angle $2\theta$ for X-Rays incident on a glass coverslip grafted with approximately 7 nm thick PDMS brush. The inset shows the plot of $\theta_{m}^2$ against $m^2$. The slope for the linear fit is 2 $\times 10^{-4}$ radians$^2$. }
	\label{xrr}
\end{figure}

To measure the thickness of the polydimethylsiloxane (PDMS) chains on glass coverslip, we employed X-ray Reflectivity (XRR) technique.  In this technique, X-rays are incident on the sample at grazing angles, and the scattered intensities from the film-air and film-solid interfaces interfere to give a resultant intensity. The reflectivity, defined as the ratio of the reflected intensity to the incident intensity, is then plotted against the scattering angle $2\theta$.
Once $2\theta>2\theta_\mathrm{c}$, where $\theta_\mathrm{c}$ is the critical angle, the oscillations in intensity (also known as Kiessig oscillations) become visible  \cite{xrr_book}. 
For a smooth thin film with thickness $h_\mathrm{film}$, analyzed at a small angle of incidence, the approximate thickness can be determined from the positions of the local maxima in the reflectivity data using the relation $\theta_{m}^2 = \theta_c^2 + \left( \lambda/2\,h_\mathrm{film} \right)^2 m^2$, where the $m$-th maximum appears at an angle $2\,\theta_{m}$.  Thus, plotting $\theta_{m}^2$ against $m^2$, yields $h_\mathrm{film} = \lambda/2\sqrt{s}$ with the slope $s$. Fig. \ref{xrr} shows the variation of reflectivity with the scattering angle $2\,\theta$ for our system.
We found the thickness of the brushes to be about 7 nm.  We also found a similar value (about 9 nm) on silicon substrates using ellipsometry (Nanofilm EP3). It is important to note that the accurate determination of the thickness requires numerical fitting of the full XRR profile. However, since the chains exhibit multilayer thickness, such numerical fitting is beyond the scope of this work \cite{pseudobrush2023}.

\clearpage
\section{Calculating the Laplace pressure of liquid bridges}
\begin{figure}[h]
	\centering
	\includegraphics[width=8.5cm]{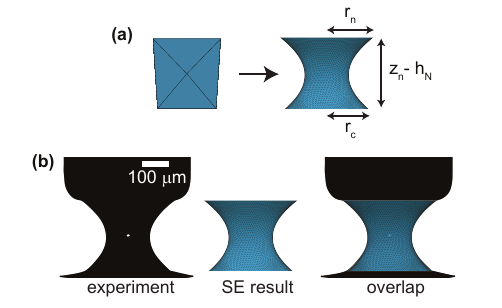}
	\caption{(a) Side view for the initial and the equilibrium configuration of Krytox on a solid surface (with same surface energy as silicone oil), held by a needle determined using Surface Evolver (SE).
		The black lines correspond to edges of the triangular mesh. The upper vertices contacting the needle are constrained to lie on a circular boundary of radius $r_\mathrm{n}$ and height 
		$z_\mathrm{n}$, while the lower vertices are constrained to lie on a circular boundary of radius $r_\mathrm{c}$. (b) The simulated result is validated by overlapping it with the experimental image.}
	\label{se_press_calculation}
\end{figure}

The Laplace pressure $\Delta P_\mathrm{liq}$ at any point on the oil-liquid interface is given by
\begin{equation}
	\Delta P_\mathrm{liq}=\Delta P_\mathrm{liq,N}+(h_\mathrm{N}-h)\rho_\mathrm{oil}\,g
\end{equation}
where  $\Delta P_\mathrm{liq,N}$ and  $h_\mathrm{N}$ corresponds to the Laplace pressure of liquid and height of thin film, respectively, at the three phase contact line. Typically, $\Delta P_\mathrm{liq,N}\gg(h_\mathrm{N}-h)\rho_\mathrm{oil}\,g$, thus, the problem of finding $\Delta P_\mathrm{liq}$ simplifies to finding $\Delta P_\mathrm{liq,N}$ due to a given liquid configuration confined between a needle and a surface (with same interfacial properties as the oil).
To address this, we defined the initial configuration of the liquid, as shown in Fig. \ref{se_press_calculation}(a). To incorporate the pinning effect due to the needle,  the upper vertices of the liquid were constrained to remain on a circular boundary with $r=r_\mathrm{n}$ and height $z=z_\mathrm{n}-h_\mathrm{N}$. Since $r_\mathrm{c}$ from the side view was known, an additional constraint was also defined for the lower vertices and edges (at $z=0$). Next, we provided the interfacial tensions $\gamma_\mathrm{LA}$ and $\gamma_\mathrm{OL}$ for the liquid-air and oil-liquid interface to the capillary surfaces, along with the liquid mass density $\rho_\mathrm{liq}$. The system's equilibrium configuration  is then determined using Surface Evolver (through a sequence of iterative steps) by minimizing the total energy with the prescribed constraints. The total energy $E$ is expressed as
\begin{equation}
	E = \gamma_\mathrm{LA} \, A_\mathrm{LA} +\Delta P_\mathrm{liq}\, V_\mathrm{liq}+ \rho_\mathrm{liq}\,g \int  z\, dV
\end{equation}
\noindent where $A_\mathrm{LA}$ is the total area of the liquid-air interface and  $\Delta P_\mathrm{liq}$ is the Lagrange multiplier for the fixed $V_\mathrm{liq}$. Physically, $\Delta P_\mathrm{liq}$ corresponds to the Laplace pressure at $z=0$. Convergence was considered to be achieved when the change in energy fell below the value of $10^{-15}$ J. The parameters $r_\mathrm{n}$ and  $z_\mathrm{n}$ were determined from the side view. The liquid volume was estimated from the spherical cap profile of liquid prior to contact. 
%To estimate the liquid volume, we calculated the volume of the spherical cap profile of liquid prior to contact. 
This estimate was then fine-tuned within the uncertainty range to achieve a good agreement between the computed and experimental results.
Fig. \ref{se_press_calculation}(b) illustrates one such example of the overlap. 
%an approximate value based on the spherical cap of the liquid held by the needle prior to contact. 

\clearpage
\section{Variation in the Laplace pressure of liquid and width of contact line}

\begin{figure}[h]
	\centering
	\includegraphics[width=8.5cm]{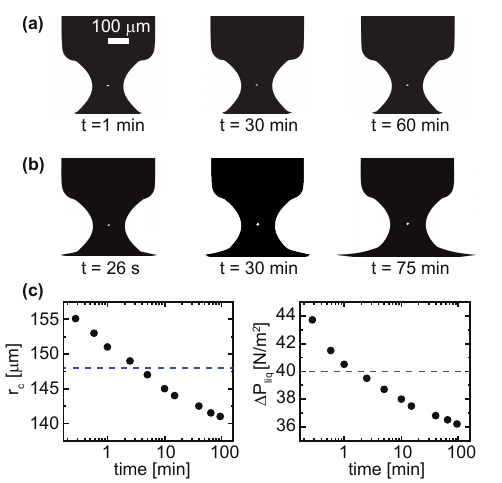}
	\caption{Variation in the Liquid's Laplace pressure and width of the three phase contact line in experiments. (a) and (b) shows the experimental side view images captured at different times after Krytox, in the form of a bridge, contacted a \SI{1.5}{\,\micro\metre} and \SI{5.9}{\,\micro\metre} thick oil film, respectively (cases $h_{0,1}$ and $h_{0,2}$ discussed in the main section). Due to faster dynamics for thicker films, the wetting ridge grows significantly faster and thus leads to a significant variation in $\Delta P_\mathrm{liq}$ and three phase contact line radius $r_\mathrm{c}$, as shown in the semi-log plot in (c). The blue horizontal dashed line corresponds to the average value ($r_\mathrm{c}= 148\SI{}{\,\micro\metre}$ and $\Delta P_\mathrm{liq} = 40\SI{}{\,\N/m^2}$) used in the numerical computations.}
	\label{exp_laplace}
\end{figure}

\clearpage

\section{Numerical scheme for studying the lubricant dynamics}
To numerically solve equation (2), we first expressed it in the dimensionless form by defining the dimensionless variables $H=h/ h_0, T=t/t_\mathrm{c}$ and $R=r/r_\mathrm{c}$ to obtain
\begin{equation}
\frac{\partial H}{\partial T}=\frac{1}{R}\frac{\partial}{\partial R}\left(R Q\right)
\end{equation}
\noindent where $Q = \sum_{i=1}^{6} T_i$ with $T_i$ defined as:
%\sum_{k=1}^{6} T_k
\begin{equation}
T_1 = \left( -\frac{\gamma_\mathrm{OL}}{\gamma_\mathrm{OA}} H^{3} (H_{RRR}-\frac{H_{R}}{{R^{2}}}+\frac{H_{RR}}{{R}}) + \frac{V_{\gamma,\text{in}} H_R}{H} \right) U(1-R)
\end{equation}
\begin{equation}
T_2=\left( \frac{H^{^3} r_\mathrm{c}^2\, \Delta P_{\mathrm{liq}}} {h_0 \,\gamma_\mathrm{OA}}-\frac{\gamma_\mathrm{OL}}{\gamma_\mathrm{OA}}H^{3} (H_{RR}+\frac{H_{R}}{R})-\frac{V_{\gamma,\mathrm{in}}}{3} \right)U_{R}(1-R)
\end{equation}
\begin{equation}
T_3=\left(-H^{3} (H_{RRR}-\frac{H_{R}}{{R^{2}}}+\frac{H_{RR}}{{R}})+\frac{V_{\gamma,\text {out}} H_{R}}{H}\right) U(R-1)
\end{equation}
\begin{equation}
T_4=\left(-H^{3} (H_{RR}+\frac{H_{R}}{R})-\frac{V_{\gamma,\text {out}}}{3} \right)U_{R}(R-1)
\end{equation}
\begin{equation}
T_5= -\frac{H^{3}\, r_\mathrm{c}}{h_0} \frac{\gamma_\mathrm{LA}}{\gamma_\mathrm{OA}} \cos \theta \,\delta_{R}(R-1)
\end{equation}
\begin{equation}
T_6= B_\mathrm{o} H_{R}H^{3}
\end{equation}
\noindent where $U$ and $\delta$ are the step and Delta functions, respectively, and  $B_\mathrm{o}, V_{\gamma,\text {in}}$, and $V_{\gamma,\text {out}}$ are defined as 
\begin{equation}
B_\mathrm{o}=\frac{\rho_\mathrm{oil} \,g \, r_\mathrm{c}^2}{\gamma_{\mathrm{OA}}}, V_{\gamma,\text {in}}=\frac{3 A_{\text {LOS }} r_\mathrm{c}^2}{\gamma_{\text{OA }} 6 \pi h_0^4},V_{\gamma,\text {out}}=\frac{3 A_{\text {AOS }} r_\mathrm{c}^2}{\gamma_{\text{OA }} 6 \pi h_0^4}
\end{equation}
\noindent To discretize the system, we adopted a similar methodology as reported by Dai and Vella \cite{vella}. 
The spatial region, $0\leq X \leq X_{\infty}$ is discretized into $N_1$ and $N_2$ cells for the regions, $0\leq X \leq 1$ and $1\leq X \leq X_{\infty}$, respectively. The value of $H$ at the mid point of the i-th cell is defined as $H_i$. The width of the i-th cell $\Delta R_i$ is distributed non-uniformly such that the cells around the three phase contact line are highly refined. The right and the left endpoints of the i-th cell are termed as $i+1/2$ and $i-1/2$, respectively. Using the central finite difference technique and method of lines, a set of ordinary differential equations (ODEs) were obtained:
\begin{equation}
    \frac{\dd H_i}{\dd T} = \frac{1}{\Delta R_i}\left(Q_{i+1/2}-Q_{i-1/2}\right) + \frac{Q_i}{R_i}
    \label{eq:ODE}
\end{equation}
where 
\begin{subequations}
\begin{align}
    Q_{i+1/2}&= \left( \sum_{k=1}^{6} T_k\right)_{i+1/2}\\
    H_{RRR}|_{i+1/2}&=2\left(H_{RR}|_{i+1}-H_{RR}|_{i}\right)/(\Delta R_i+\Delta R_{i+1})\\
    H_{RR}|_{i}&=\left(H_R|_{i+1/2}-H_R|_{i-1/2}\right)/\Delta R_i\\
    H_R|_{i+1/2}&=2(H_{i+1}-H_i)/(\Delta R_{i+1}+\Delta R_i)\\
    H_{i+1/2}&=\left(\Delta R_{i+1}H_i+\Delta R_i H_{i+1}\right)/(\Delta R_{i+1}+\Delta R_i)
\end{align}
\end{subequations}
\noindent To connect the two menisci automatically at the three phase contact line, we used the smoothed forms of the unit step functions and dirac delta functions, expressed as:
\begin{subequations}
\begin{align}
\delta(R-1)&=\frac{1}{2\epsilon} \operatorname{sech}^2\left( \frac{1-R}{\epsilon}\right) \\
U(1-R)&=\frac{1}{2} \left(1+\tanh\frac{1-R}{\epsilon}\right)\\
U(R-1)&=\frac{1}{2} \left(1+\tanh\frac{R-1}{\epsilon}\right)
\end{align}
\end{subequations}
\noindent where $\epsilon\to0$. The derivatives thus simplify to
$\delta_{R}(R-1)=\frac{1}{\epsilon^2} \operatorname{sech}^2(\frac{1-R}{\epsilon})\tanh(\frac{1-R}{\epsilon})$, $U_{R}(1-R)=-\frac{1}{2\epsilon} \operatorname{sech}^2(\frac{1-R}{\epsilon}),$ and $U_{R}(R-1)=\frac{1}{2\epsilon} \operatorname{sech}^2(\frac{R-1}{\epsilon})$. The ODEs were then solved using MATLAB's builtin solver ode15s, with the folowing boundary conditions: 
\begin{equation}
    H_R(0,T)=H_R(R_{\infty},T)=H_{RRR}(0,T)=H_{RRR}(R_{\infty},T)=0
\end{equation}

\newpage
\section{Variation in maximum height at the center in stage-I and waiting period in stage-III}
\begin{figure}[h]
    \centering
    \includegraphics[width=17.8cm]{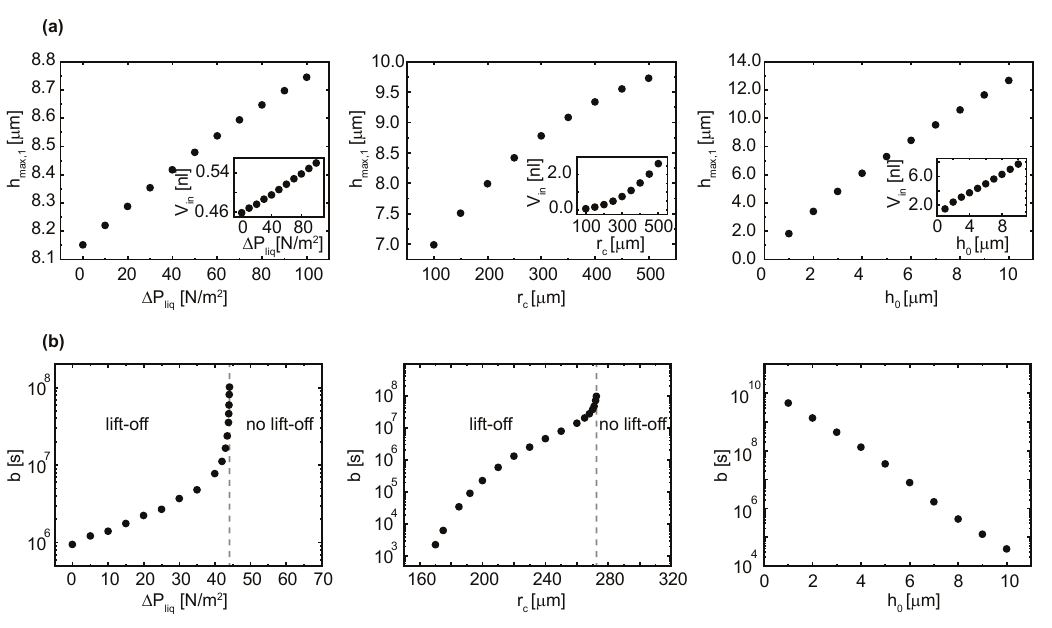}
    \caption{Variation in maximum height at the center in stage-I and waiting period in stage-III with system parameters. (a) Variation in maximum height reached by the center in stage-I, $h_\mathrm{max,1}$ with system parameters: initial film thickness $h_0$, Laplace pressure of Krytox bridge $\Delta P_\mathrm{liq}$, and width of three phase contact line $r_\mathrm{c}$. While varying one parameter, the other two parameters are kept fixed. The values of the fixed parameters are $\Delta P_\mathrm{liq} = 40$ N/m$^2$, $h_\mathrm{0} = 6\SI{}{\,\micro\metre}$, and $r_\mathrm{c}=$ \SI{250}{\,\micro\metre}.  $h_\mathrm{max,1}$ increases non-monotonically with the system parameters. The reason is the non-monotonic increase in oil volume $V_\mathrm{in}$ beneath Krytox with system parameters as shown in the respective insets.  $V_\mathrm{in}$ is calculated using $V_\mathrm{in} = \int_{r=0}^{r_\mathrm{edge}} 2\pi r h dr$ where $r_\mathrm{edge}$ is the position of the edge from the center. (b) Variation of waiting time period $b$ in stage-III with system parameters. While $b$ increases non monotonically with $\Delta P_\mathrm{liq}$ and $r_\mathrm{c}$, it decreases with $h_0$. By changing the value of $\Delta P_\mathrm{liq}$, and $r_\mathrm{c}$, the liquid configuration can be tuned to lie in the lift-off or no lift-off configuration.    
    }
    \label{numerical_predictions_for_b_and_hmax}
\end{figure}
\clearpage

\section{Variation in the dimensionless position and height over time}

\begin{figure}[h]
	\centering
	\includegraphics[width=17.8cm]{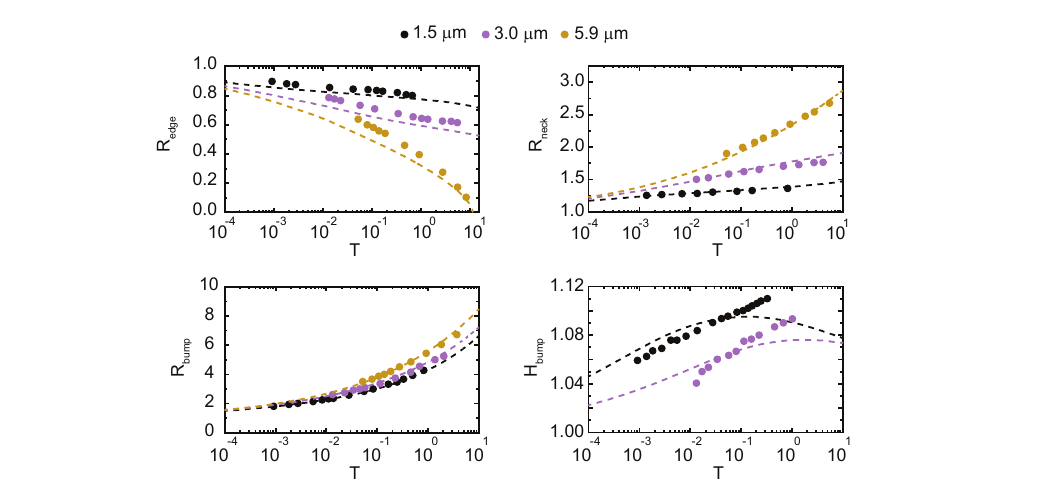}
	\caption{Change in dimensionless position and height over time after Krytox in the form of a liquid bridge contacts a silicone oil coated glass. The results are illustrated for the three cases $h_{0,1},h_{0,2}$, and $h_{0,3}$ with initial thicknesses 
	$h_{0}=1.5 , 3.0$, and $5.9$ \SI{}{\,\micro\metre}, respectively (see main section for details). The dimensionless positions of edge, neck and bump are represented by $R_\mathrm{edge}$, $R_\mathrm{neck}$, and $R_\mathrm{bump}$, respectively, while the dimensionless height of the bump is represented by $H_\mathrm{bump}$. The filled circles correspond to the experiments and the dashed lines (with the same color as the corresponding experiment) represent the numerical predictions. 
	}
	\label{bump_edge_neck_suppl}
\end{figure}

\newpage
\section{Determining the equilibrium configuration}
\begin{figure}[h]
    \centering
    \includegraphics[width=8.5cm]{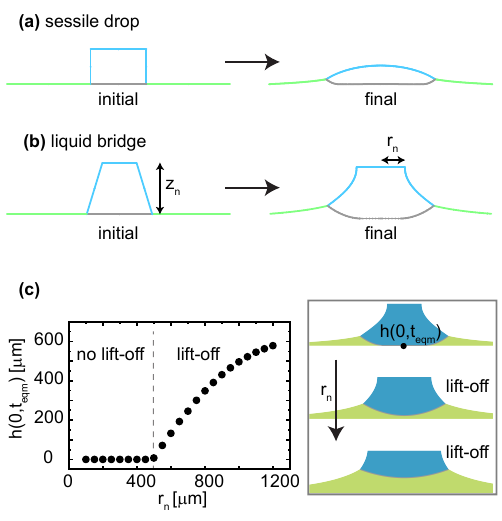}
    \caption{Equilibrium configuration of system computed using Surface Evolver. (a) and (b) side slice views for initial and equilibrium configuration of Krytox in the form of sessile droplet and liquid bridge, respectively, on a silicone oil coated glass. For the bridge configuration, the upper vertices of Krytox in contact with the needle are constrained to lie on a circular boundary with radius $r_\mathrm{n}$ and height $z_\mathrm{n}$, mimicking the effect of the needle. (c) Variation in equilibrium film height at the center  $h(0,t_\mathrm{eqm})$ as a function of $r_\mathrm{n}$ for fixed $z_\mathrm{n}=\SI{850}{\micro\metre}$ and liquid bridge with $V_\mathrm{L}= 1 \SI{}{\micro\litre}$. For $r_\mathrm{n}>500\SI{}{\,\micro\metre}$, the edge merge and the liquid lift-off occurs. The equilibrium configurations for $r_\mathrm{n}= 350, 550, \textrm{and } 750\SI{}{\,\micro\metre}$, arranged from top to bottom, are displayed in the right panel.}   
    \label{eqm_se}
\end{figure}

The equilibrium configuration of system was numerically computed using Surface Evolver. The initial configuration for the oil and liquid was defined, as shown in Fig. \ref{eqm_se}(a),(b). The oil and liquid were defined as two separate bodies with their mass densities $\rho_\mathrm{oil}$ and $\rho_\mathrm{liq}$, respectively. The capillary surfaces representing the liquid-air, oil-air, and oil-liquid interfaces were assigned interfacial tensions $\gamma_\mathrm{LA}$, $\gamma_\mathrm{OA}$, and $\gamma_\mathrm{OL}$, respectively.
For the oil, Laplace pressure at $z=0$ was set to zero, while for the liquid, a constant volume constraint was applied. To incorporate the pinning effect caused by a needle for the case of a liquid bridge, the upper vertices of the liquid were constrained to lie on a circular boundary with $r=r_\mathrm{n}$ and at height $z=z_\mathrm{n}$. The system's equilibrium configuration was then determined using Surface Evolver by minimizing the total energy $E$, expressed as  \cite{SE, surfaceevolvermanual, semprebon_2017,shivam_2023}
\begin{equation}
	E = \sum_{i \neq j} \gamma_{ij} \, A_{ij}+
	\Delta P_\mathrm{liq}\, V_\mathrm{liq}+ +\Delta P_\mathrm{oil} V_\mathrm{oil}+\rho_\mathrm{liq}\,  g\int_\mathrm{liq}  z \, dV+\rho_\mathrm{oil}\,  g\int_\mathrm{oil}  z \, dV
\end{equation}
\noindent where $i,j$ correspond to fluid-fluid interface with total interface area $A_{ij}$. $\Delta P_\mathrm{liq}$ and $\Delta P_\mathrm{oil}$ are the Lagrange multipliers for the liquid and oil volume, $V_\mathrm{liq}$ and $V_\mathrm{oil}$, respectively. Convergence was considered to be achieved when the free energy change fell below the value of $10^{-15}$ J. To examine the effect of different  $\Delta P_\mathrm{liq}$, the configurations were computed for different $z_\mathrm{n}$ and $r_\mathrm{n}$ for the case of liquid bridge and different $V_\mathrm{liq}$ for the case of sessile droplet. 

\newpage

\clearpage
\section{Effect of initial film thickness and number of droplets on the size of wetting ridge}
\begin{figure}[h]
    \centering
    \includegraphics[width=17.8cm]{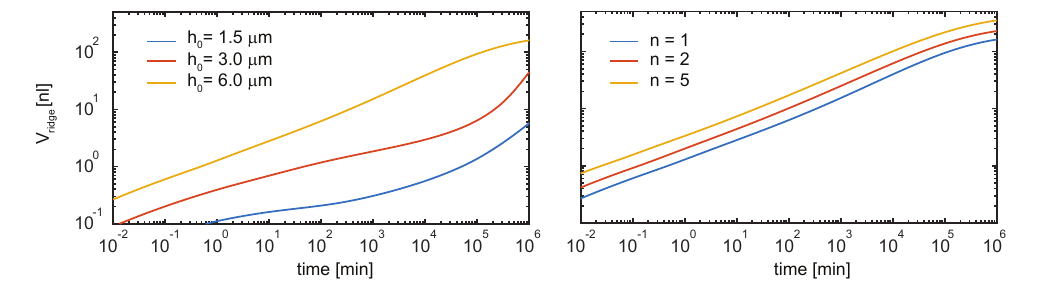}
    \caption{Temporal variation of wetting ridge volume $V_\mathrm{ridge}$ for a Krytox droplet. The total droplet volume $V$ corresponding to a spherical cap with $r_\mathrm{c}=$ \SI{250}{\,\micro\metre} and $\theta=46^\circ$ is kept constant. Left: Variation of $V_\mathrm{ridge}$ for different initial film thicknesses $h_\mathrm{0}$ for a droplet of volume $V$. The plot concludes that thicker films result in a larger wetting ridge.  Right: Variation of $V_\mathrm{ridge}$ for different number of identical droplets ($n$) that collectively sum to the total volume $V$. The intial thickness is $h_0=6.0\SI{}{\,\micro\metre}$. The plot concludes that multiple smaller droplets result in larger wetting ridge as compared to a single large droplet with same volume. For the calculation, we assumed that the film dynamics caused by each droplet is independent and unaffected by the other droplets.
    }
    \label{}
\end{figure}

\clearpage
\bibliography{references}% common bib file